\documentclass[conference]{IEEEtran}
\usepackage[T1]{fontenc}
\usepackage[latin9]{inputenc}
\usepackage{array}
\usepackage{mathtools}
\usepackage{amsmath,amssymb}
\usepackage{graphicx}
\usepackage{stfloats}
\usepackage{float}
\newcommand\scalemath[2]{\scalebox{#1}{\mbox{\ensuremath{\displaystyle #2}}}}
\usepackage[unicode=true,
 bookmarks=true,bookmarksnumbered=true,bookmarksopen=true,bookmarksopenlevel=1,
 breaklinks=false,pdfborder={0 0 1},backref=false,colorlinks=false]
 {hyperref}
\hypersetup{pdftitle={Your Title},
 pdfauthor={Your Name},
 pdfborderstyle=,pdfpagelayout=OneColumn,pdfnewwindow=true,pdfstartview=XYZ,plainpages=false}

\makeatletter

\providecommand{\tabularnewline}{\\}
\makeatother

\begin{document}
\title{An Optimized Nearest Neighbor Compliant Quantum Circuit for 5-qubit Code}
\author{Arijit Mondal and Keshab K. Parhi, {\em Fellow, IEEE}\\
 Email: \{monda109, parhi\}@umn.edu\\
 Department of Electrical and Computer Engineering, University of
Minnesota}
\maketitle
\begin{abstract}
The five-qubit quantum error correcting code encodes one logical qubit to five physical qubits, and protects the code from a single error. It was one of the first quantum codes to be invented, and various encoding circuits have been proposed for it. In this paper, we propose a systematic procedure for optimization of encoder circuits for stabilizer codes. We start with the systematic construction of an encoder for a five-qubit code, and optimize the circuit in terms of the number of quantum gates. Our method is also applicable to larger stabilizer codes. We further propose nearest neighbor compliant (NNC) circuits for the proposed encoder using a single swap gate, as compared to three swap gates in a prior design.
\end{abstract}


\section{Introduction}

Quantum computing has the potential to revolutionize the state-of-the-art computing and communications technologies. Some prior advances in quantum computing include:  Shor's factorization \cite{shor1994algorithms} and Grover's search algorithm \cite{grover1996fast}. Realizations of these algorithms in real life require large quantum computers. Although current quantum computers are not large enough to solve practical problems, we can expect such computers to be available in near future. However, packing a large number of qubits leads to more errors due to noise and decoherence. Thus, we need mechanisms that can be coupled with quantum computers to mitigate errors introduced by noise.

Quantum error correction was believed to be impossible till Shor proposed a 9-qubit code with the ability to correct a single quantum error of any type \cite{shor}. Calderbank, Shor, and Steane proposed a method, referred to as CSS framework, to import the vast literature of classical error correction codes (ECC) to construct new quantum ECCs \cite{calderbank,steane1996multiple}. Subsequently, Gottesman developed a stabilizer framework for quantum ECCs which became widely popular \cite{gottesman-thesis}. Entanglement assisted (EA) stabilizer codes were proposed in \cite{brun2006correcting}, leading to further increase in the error correction ability \cite{lai2013entanglement}. Subsequently, classical ECCs like low-density parity-check (LDPC), polar codes, and Reed-Solomon (RS) codes were constructed in the quantum domain \cite{hsieh2009entanglement,dupuis2019purely,grassl1999quantum,aly2008asymmetric,la2012asymmetric}. 

Encoding and decoding circuits for qubit stabilizer codes were proposed by Gottesman \cite{gottesman-thesis}. An approach towards encoder design that was applicable to qudits was presented in \cite{grassl2003efficient}. These circuits used different types of 2-qubit gates, such as CNOT, CY, and CZ gates along with single qubit Pauli gates. For practical circuit realization, it is often useful to design a circuit using a single type of 2-qubit gate. Also, the circuits proposed after systematic design in \cite{gottesman-thesis} can be further optimized in terms of the number of gates. For a review of quantum error correction using stabilizer codes, the reader is referred to \cite{mondal2023quantum}.

CNOT gates along with other single-qubit Pauli gates form a universal set of gates, i.e., they can be used to generate any arbitrary unitary transformation for a $n$-qubit system. A set of equivalencies exist between quantum gates, similar to classical digital gates. Such equivalencies were used to study quantum circuits in \cite{zhou2000methodology,mermin2001classical,mermin2002deconstructing,maslov2008quantum}. These rules, along with some new formulated rules were compiled into a set of equivalence rules in \cite{garcia}. To enhance circuit reliability, it is necessary to optimize the quantum circuits in terms of the number of CNOT gates. Following the works in \cite{alber2001quantum,patel2008optimal}, the authors in \cite{bataille2022quantum} studied the one-to-one correspondence between $n$-qubit systems consisting of CNOT gates and $n\times n$ non-singular matrices with coefficients in $\mathbb{F}_2$, in a formal way using group isomorphism and group representation. Using this approach, they proposed an algorithm to optimize a system of $n$ qubits, with CNOT gates interacting between them.

Various quantum technologies offer different degrees of freedom in terms of interaction between qubits \cite{imany2019high,liu2018coupled,klimov2003qutrit,leon2020coherent}. Non adjacent qubit interactions are heavily prone to noise. Many of the fault-tolerant technologies thus rely on the nearest neighbor compliance, where the interacting qubits need to be adjacent to each other before they can interact via 2-qubit gates. Various techniques have been proposed in the past to design 1-D and 2-D architectures that are nearest neighbor compliant (NNC) \cite{saeedi2011synthesis,chakrabarti2011linear,matsuo2012changing,shafaei2013optimization,markov2012constant,pham20132d,shrivastwa2015fast,alfailakawi2016harmony,lin2014paqcs}.

For both 1-D and 2-D quantum architectures, swap gates are used to move qubits adjacent to each other before performing an operation between them. A swap gate can be realized by 3 CNOT gates. Irrespective of the quantum technology or quantum architecture used, minimizing the number of swap gates is of prime importance towards design of efficient quantum circuits. Approaches towards getting the lowest number of swap gates for a quantum circuit were studied in \cite{alfailakawi2016harmony,lye2015determining,ding2019exact}.  Quantum ECC circuits should be designed subject to NNC constraint. This is extremely important for fault-tolerant computing. Without NNC constraint, the error correcting circuits will be prone to noise and errors, and may introduce new errors while correcting the existing ones.

The five-qubit code is the smallest quantum ECC which can correct a single qubit error \cite{bennett,laflamme}. Gottesman \cite{gottesman-thesis} studied the 5-qubit code using stabilizer formalism. Authors in \cite{gong2022experimental} experimentally verified a five-qubit code encoder using six 2-qubit gates and twelve single qubit gates. They derived the circuit through manual search, but further details were not provided. We believe that the manual search may imply searching for optimal circuit among clifford group elements for 5 qubits \cite{bravyi20226}. However, such a method may not be possible for larger quantum ECCs since the size of the clifford group increases significantly. In this paper our goal is to design an optimal circuit in a systematic way starting from the circuit in \cite{gottesman-thesis}.

The contributions of this paper are as follows. First, we design a 5-qubit encoder circuits using only CNOT gates and Hadamard gates, $H$. Second, we propose a systematic method to optimize encoder circuit for stabilizer codes in terms of the number of 2-qubit gates.  Third, for a two-dimensional (2-D) qubit configuration, our optimized encoder requires only one swap gate for nearest neighbor compliance compared to the initial circuit \cite{gottesman-thesis} which requires 3 swap gates.
%We simulate and test all the proposed circuits using IBM Qiskit.

The rest of the paper is organized as follows. In Section II, we give a brief theoretical background followed by the systematic construction of the encoder. In Section III, we provide details of the optimization of the encoder circuit. We provide NNC circuits for the encoder in Section IV, followed by results in Section V. 
%We conclude the paper in Section VI.

%\section{Theoretical background}

\section{Theoretical background and encoder design for 5-qubit code}

A set of qubits which evolve through a sequence of unitary operations constitute a quantum circuit. The
unitary operations are represented by quantum gates. The single qubit gates include: bit flip gate
$X$, phase flip gate $Z$, Hadamard gate $H$, $Y$ gate, and the
phase gate $S$. 
The multi-qubit gates include: controlled-$X$ (CNOT), controlled-$Z$ (CZ), controlled-$Y$ gates, and the CCNOT (Toffoli gate). For matrix representations of the gates and further reading on quantum circuits, the reader is referred to \cite{nielsen}.

The five-qubit code can correct a single qubit error \cite{bennett,laflamme}. Gottesman \cite{gottesman-thesis} studied the 5-qubit code using the stabilizer formalism. The stabilizers $M_1-M_4$  and logical $\bar{X}$ and $\bar{Z}$ operators for a 5-qubit ECC are shown below:
\begin{center}
\begin{tabular}{c|ccccc}
$M_{1}$ & $X$ & $Z$ & $Z$ & $X$ & $I$\tabularnewline
$M_{2}$ & $I$ & $X$ & $Z$ & $Z$ & $X$\tabularnewline
$M_{3}$ & $X$ & $I$ & $X$ & $Z$ & $Z$\tabularnewline
$M_{4}$ & $Z$ & $X$ & $I$ & $X$ & $Z$\tabularnewline
$\bar{X}$ & $X$ & $X$ & $X$ & $X$ & $X$\tabularnewline
$\bar{Z}$ & $Z$ & $Z$ & $Z$ & $Z$ & $Z$\tabularnewline
\end{tabular}
\par\end{center}

% The basis codewords can be written as

% \begin{equation}
% |\bar{0}\rangle=\sum_{M\in S}M|00000\rangle
% \end{equation}

% \begin{equation}
% |\bar{1}\rangle=\bar{X}|\bar{0}\rangle
% \end{equation} 

\begin{figure}
\begin{centering}
\includegraphics[scale=0.34]{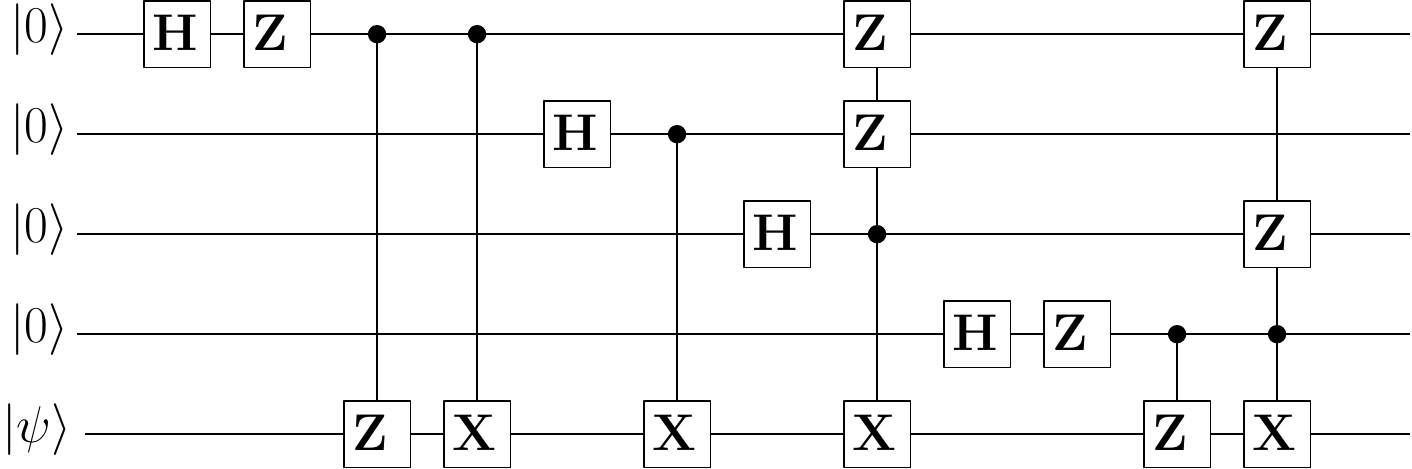}
\par\end{centering}
\caption{Encoder for the five-qubit code using CNOT and CZ as 2-qubit gates \cite{gottesman-thesis,mondal2023quantum}. \label{fig:five-qubit-encoder-XZ}}
\end{figure}

Using Gottesman's method \cite{gottesman-thesis}, an algorithm for designing an encoding circuit was presented in \cite{mondaltcas12024}. Using the algorithm, the encoding circuit for the 5-qubit code is shown in Fig.
\ref{fig:five-qubit-encoder-XZ}. We don't go into the details of this procedure since our contribution is towards the optimization and developing NNC circuit for the encoder. Interested readers are referred to \cite{gottesman-thesis,mondal2023quantum} for more details.

\section{Optimization of the encoder circuit}

We observe from Fig. \ref{fig:five-qubit-encoder-XZ} that the circuit requires 4 $H$ gates, 2 $Z$ gates, 6 CZ gates and 4 CNOT gates \footnote{It should be noted that if CY gates are also used, a total of 8 2-qubit gates are required. However, a CY gate is equivalent to two gates, as it is a combination of CNOT and CZ gate}. 2-qubit gates are more noisy than single qubit gates. To mitigate noise, our goal is to optimize the circuit in terms of the number of gates. Also, for practical implementations, it is often necessary to design the circuit using one type of 2-qubit gate. Thus, our objective is to design a circuit which contains only CNOT gates and single-qubit Pauli gates.

\subsection{Optimization using equivalence rules}

In \cite{garcia}, the authors discuss a set of equivalence rules related to quantum circuits. These were compiled into a list of 10 equivalence rules in \cite{mondaltcas12024}. We do not include the figure due to lack of space. The reader is referred to Fig. 5 in \cite{mondaltcas12024}. Rule 1 allows the conversion of $X$ gate to $Z$ gate and vice versa. Similarly, Rule 3 facilitates the conversion between CNOT and CZ gates. Rule 2 implies that the target and control can be interchanged for CZ gates. For CNOT gates, control and target can be switched with additional $H$ gates as shown in Rule 7. Rule 4 shows that a $|0\rangle$ at the control of a CNOT gate leads to no change in the output state. Similarly, a $|+\rangle$ at the target of a CNOT gate leaves the output state unaltered as shown in Rule 5. Rule 9 shows that CNOT gates commute if they share common controls or targets. When CNOT gates don't share controls or target, Rule 6 becomes helpful. Rule 8 becomes particularly useful when there is a CNOT gate between two non-adjacent qubits. A third qubit which is adjacent to both the qubits is used as an intermediate qubit for the operation. Rule 10 is a consequence of Rule 8.

% \begin{figure*}
% \begin{centering}
% \includegraphics[scale=0.34]{Equivalence-rules.pdf}
% \par\end{centering}
% \centering{}\caption{Equivalence rules related to quantum circuits (based on \cite{garcia}).}
% \label{fig:eq-rules}
% \end{figure*}

\subsection{Optimization using matrix equivalence}

A one-to-one correspondence between $n$-qubit CNOT circuits and $n \times n$ non-singular matrices was formally described in \cite{bataille2022quantum}. Let us consider a $n$-qubit circuit consisting of only CNOT gates. We can represent the initial state as a $n \times n$ identity matrix. Each CNOT gate then represents an elementary row transformation. For example, a CNOT gate with control on qubit $n_1$ and target at qubit $n_2$ can be represented by the elementary row transformation $R_{n_2}\rightarrow R_{n_1}+R_{n_2}$. Thus, the circuit can be represented by a series of elementary row transformations. We will illustrate this using an example. Let us consider rule 10 (Fig. 5 in \cite{mondaltcas12024}). The first circuit can be represented by the elementary row transformations $R_2\rightarrow R_2+R_1$ and $R_3\rightarrow R_3+R_1$ as shown below.

\[
\scalemath{0.9}{
\left[\begin{array}{ccc}
1 & 0 & 0\\
0 & 1 & 0\\
0 & 0 & 1
\end{array}\right]\rightarrow\left[\begin{array}{ccc}
1 & 0 & 0\\
1 & 1 & 0\\
0 & 0 & 1
\end{array}\right]\rightarrow\left[\begin{array}{ccc}
1 & 0 & 0\\
1 & 1 & 0\\
1 & 0 & 1
\end{array}\right]
}
\]

The second circuit can be represented by the elementary row transformations $R_3\rightarrow R_3+R_2$, $R_2\rightarrow R_2+R_1$ and $R_3\rightarrow R_3+R_2$ as shown below.

\[
\scalemath{0.9}{
\left[\begin{array}{ccc}
1 & 0 & 0\\
0 & 1 & 0\\
0 & 0 & 1
\end{array}\right]\rightarrow\left[\begin{array}{ccc}
1 & 0 & 0\\
0 & 1 & 0\\
0 & 1 & 1
\end{array}\right]\rightarrow\left[\begin{array}{ccc}
1 & 0 & 0\\
1 & 1 & 0\\
0 & 1 & 1
\end{array}\right]\rightarrow\left[\begin{array}
{ccc}
1 & 0 & 0\\
1 & 1 & 0\\
1 & 0 & 1
\end{array}\right]
}
\] 

We observe that the final matrix in both the cases are the same, and thus the circuits are equivalent. As mentioned before, the authors in \cite{bataille2022quantum} observed that for a $n$-qubit quantum circuit consisting of only CNOT gates, the input-output transformation of states can be represented by a series of elementary row transformations. Let the matrix obtained through the series of row transformations be $T_n$. Applying a Gaussian elimination method to the $T_n$ results in the identity matrix. If we apply those Gaussian elimination steps in reverse to the identity matrix, it results in $T_n$. These steps lead to an equivalent circuit that can be used as a starting point. However, this method may not always lead to an optimum circuit \cite{bataille2022quantum}. Our goal is to find the minimum number of elementary row transformations which takes the identity matrix to the matrix $T_n$. Finding an optimal solution is known to be difficult, and the complexity scales exponentially with the number of qubits involved. 

\subsection{Optimization of the 5-qubit encoder circuit}

\begin{figure}[H] 
\begin{centering}
\includegraphics[scale=0.29]{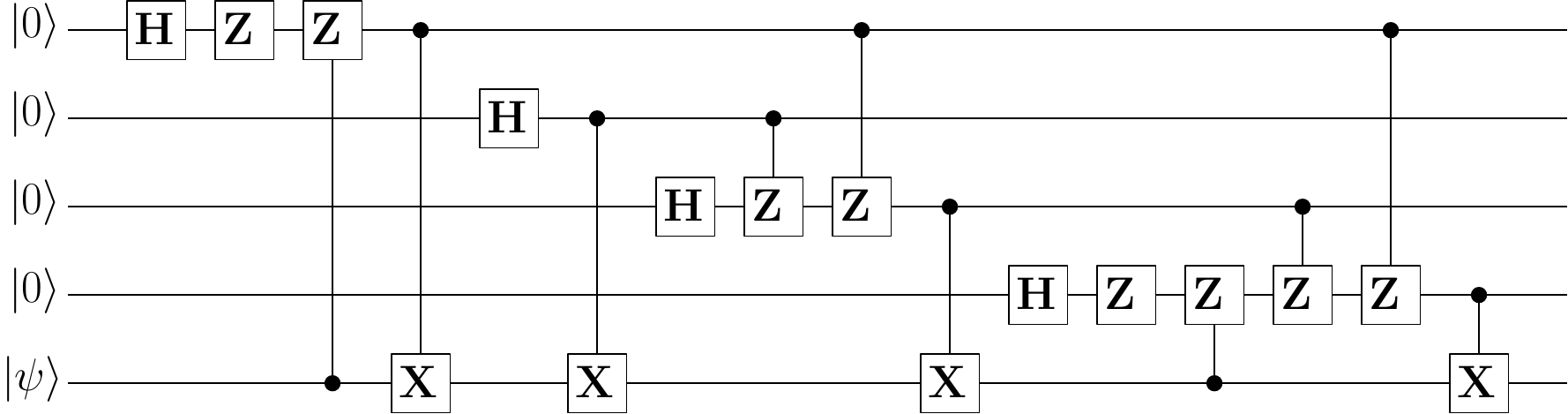}
\par\end{centering}
\centering{}\caption{Swapping control and targets for the CZ gates in the encoder in Fig. \ref{fig:five-qubit-encoder-XZ}.}
\label{fig:five-qubit-encoder-Z-reversal}
\end{figure}

The encoder circuit for the 5-qubit code using CNOT and CZ gates as two-qubit gates is shown in Fig. \ref{fig:five-qubit-encoder-XZ}. 
% First we swap the CNOT and CZ gates in the green shaded regions. This will lead to removal of two $Z$ gates after the two $H$ gates. 
Our first step would be to convert the $Z$ and CZ gates to $X$ and CNOT gates, respectively, such that there is only one type of two-qubit gate in the circuit. However, applying this to the circuit in Fig. \ref{fig:five-qubit-encoder-XZ} leads to an increase in the number of $H$ gates. To solve this issue, we first swap control and target for all the CZ gates as shown in Fig. \ref{fig:five-qubit-encoder-Z-reversal}.

\begin{figure}[H] 
\begin{centering}
\includegraphics[scale=0.31]{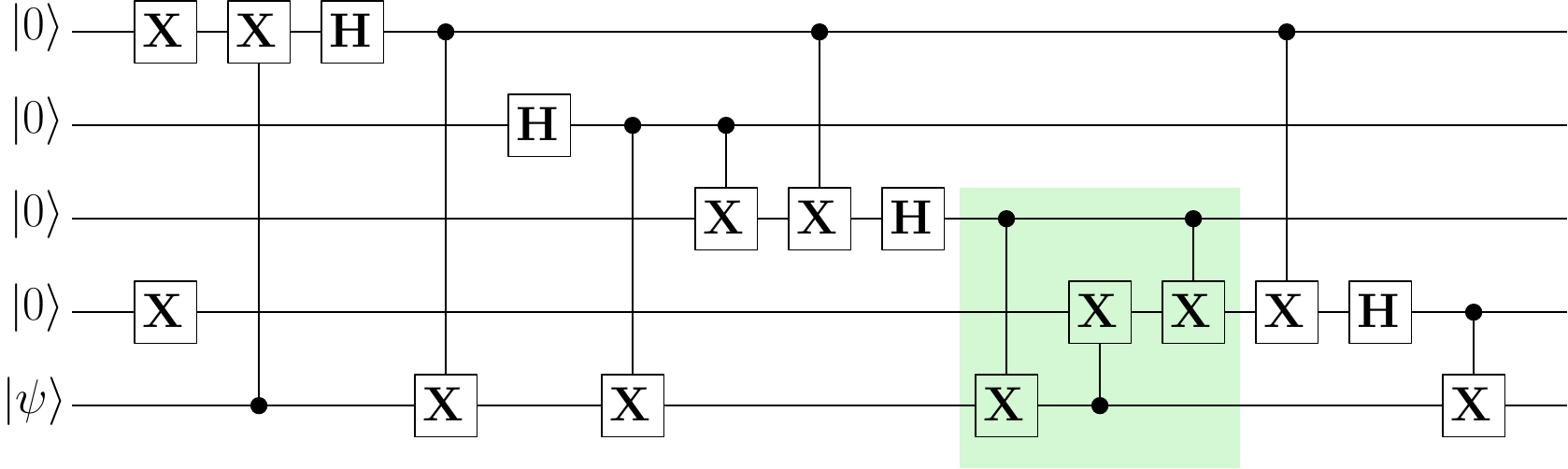}
\par\end{centering}
\centering{}\caption{Resulting encoder circuit after removing redundant $H$ gates.}
\label{fig:five-qubit-encoder-H-removal}
\end{figure}

For the next step, we convert the $Z$ and CZ gates to $X$ and CNOT gates, respectively, using Rule 3. Each CZ gate introduces two $H$ gates when converted to CNOT gates. The same is true for $Z$ gates. However, two $H$ gates are also annihilated in the process, leaving the count of $H$ gates the same. The resulting circuit after the gate conversions is shown in Fig. \ref{fig:five-qubit-encoder-H-removal}.

\begin{figure}[H] 
\begin{centering}
\includegraphics[scale=0.37]{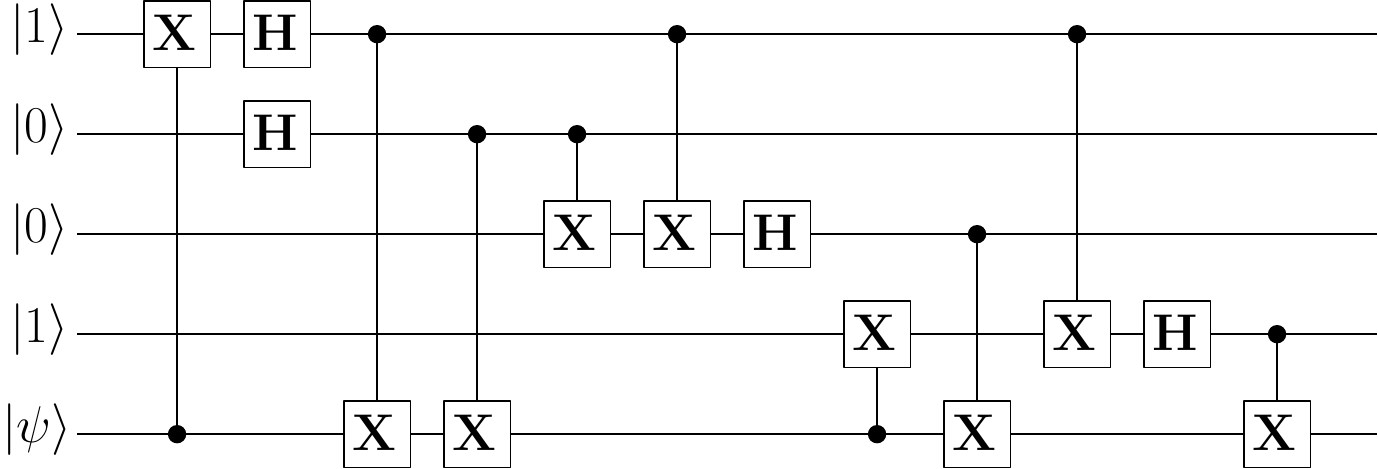}
\par\end{centering}
\centering{}\caption{Resulting encoder circuit after using rule 6 to the shaded region in Fig. \ref{fig:five-qubit-encoder-H-removal}}
\label{fig:five-qubit-encoder-opt-1}
\end{figure}

Using Rule 6, i.e., CNOT mirroring, the shaded green region in Fig. \ref{fig:five-qubit-encoder-H-removal} can be reduced from 3 CNOT gates to 2 CNOT gates. Also, the two $X$ gates on the first and fourth qubits can be removed by initializing those qubits to $|1\rangle$. The resulting circuit is shown in Fig. \ref{fig:five-qubit-encoder-opt-1}.

\begin{figure}[H] 
\begin{centering}
\includegraphics[scale=0.37]{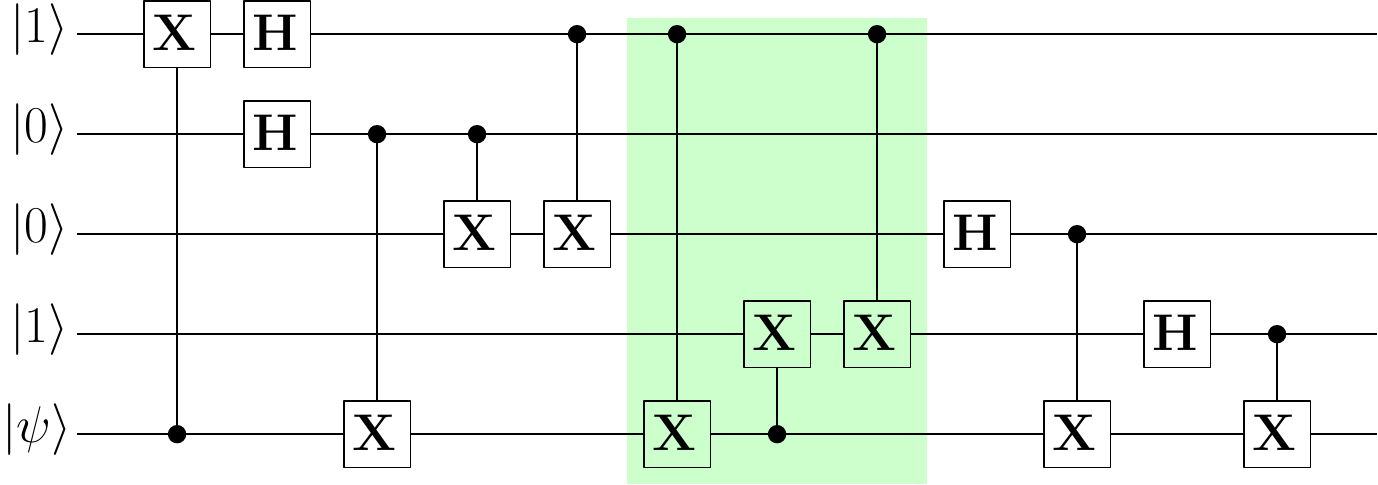}
\par\end{centering}
\centering{}\caption{Resulting encoder circuit after rearranging the circuit in Fig. \ref{fig:five-qubit-encoder-opt-1}}
\label{fig:five-qubit-encoder-opt-2}
\end{figure}

After slight rearrangement of the circuit in Fig. \ref{fig:five-qubit-encoder-opt-1}, we obtain the circuit in Fig. \ref{fig:five-qubit-encoder-opt-2}.

\begin{figure}[H] 
\begin{centering}
\includegraphics[scale=0.4]{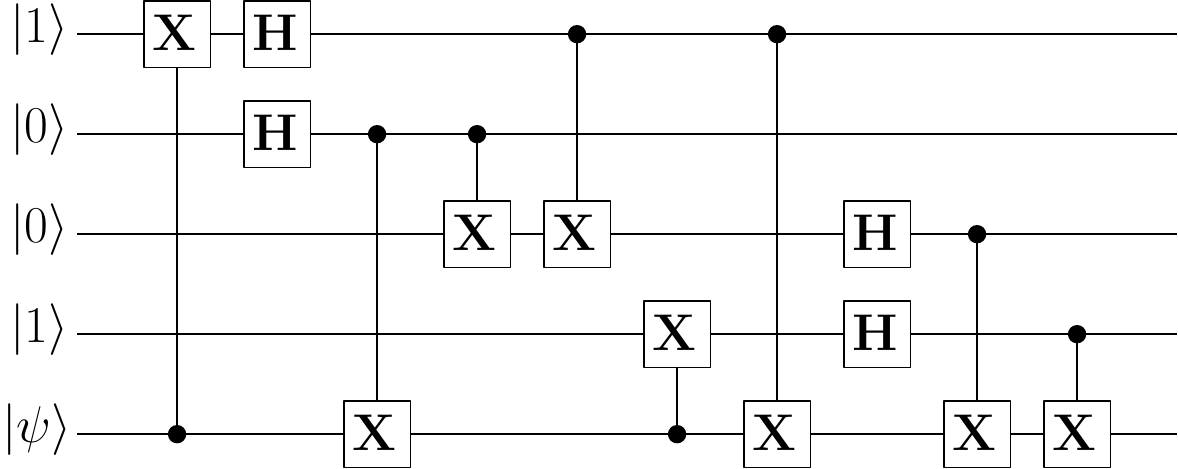}
\par\end{centering}
\centering{}\caption{Final optimized encoder circuit for the five-qubit code.}
\label{fig:five-qubit-encoder-opt-3}
\end{figure}

The green shaded region in Fig. \ref{fig:five-qubit-encoder-opt-2} can be reduced from 3 CNOT gates to 2 CNOT gates using Rule 6, resulting in the final optimized circuit as shown in Fig. \ref{fig:five-qubit-encoder-opt-3}. This circuit requires 8
CNOT gates and 4 H gates.

\section{Nearest neighbor compliant encoder circuit for the 5-qubit code}

\begin{figure}
\begin{centering}
\includegraphics[scale=0.55]{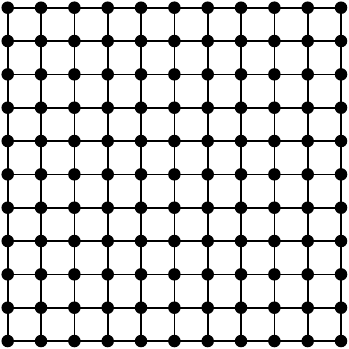} 
\par\end{centering}
\caption{2-D array of qubits (represented by black dots). \label{fig:2D-array}}
\end{figure}

\begin{figure}
\begin{centering}
\includegraphics[scale=0.5]{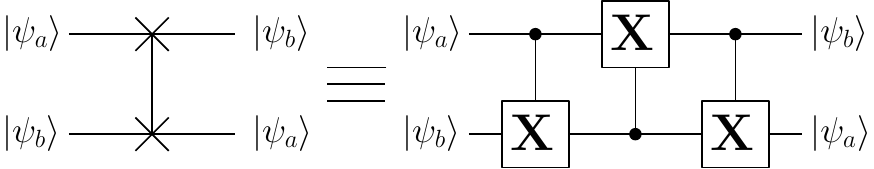} 
\par\end{centering}
\caption{Symbol of a swap gate (left). A swap gate circuit implemented using 3 CNOT gates (right). \label{fig:swap-gate}}
\end{figure}

Let us consider a 2-D array of qubits as shown in Fig. \ref{fig:2D-array}. The qubits at the corners and edges can interact with 2 or 3 qubits, while the rest of the qubits can interact with their 4 closest neighbors. To design a circuit which is nearest neighbor compliant, we need to use swap gates to bring the qubits adjacent to each other \cite{ding2019exact}. It should be noted that the qubits are not moved physically. Their states are swapped which is equivalent to moving them to adjacent positions without doing  it physically. A swap gate requires 3 CNOT gates. Thus, it is important to position the qubits and perform the operations in such a way that the number of swap gates is minimized. A swap gate is shown in Fig. \ref{fig:swap-gate}.

\begin{figure*}[b]
\begin{centering}
\includegraphics[scale=0.3]{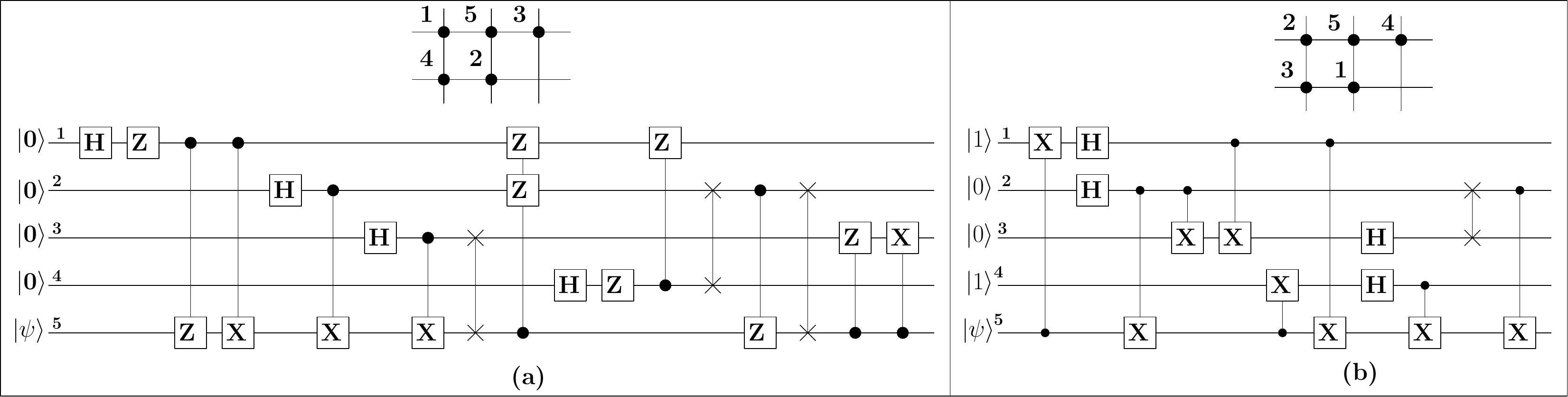}
\par\end{centering}
\centering{}\caption{(a) NNC circuit for the 5-qubit encoder without optimization, requiring 3 swap gates. (b) NNC circuit for the optimized 5-qubit encoder requiring 1 swap gate.}
\label{fig:5-qubit-NNC-opt}
\end{figure*}

We designed an NNC circuit for the 5-qubit encoder in Fig. \ref{fig:five-qubit-encoder-XZ}. From analyzing the circuit, we observe that $3$ swap gates are required to design a NNC circuit for a 2-D qubit configuration as shown in Fig \ref{fig:5-qubit-NNC-opt} (a). However, the proposed optimized circuit in Fig. \ref{fig:five-qubit-encoder-opt-3} requires only 1 swap gate for the qubit configuration shown in Fig. \ref{fig:5-qubit-NNC-opt} (b).

\section{Results}

The encoder circuit was simulated and verified using IBM Qiskit. 
%To test for correctability, a syndrome measurement circuit was also designed. Various single qubit errors were introduced at different positions to validate the correctness of the algorithm. The initial circuit, the fully optimized circuit, along with all the intermediate circuits were tested and verified. 
A table containing the resource utilization in terms of the number of gates is shown in Table \ref{tab:Resource-utilization}. We observe that, compared to the initial encoder circuit, the optimized circuit requires 4 less gates. Also, the NNC circuit for the optimized encoder requires only 1 swap gate, leading to a saving of 6 CNOT gates. The circuit in \cite{gong2022experimental} requires 6 CNOT gates and 12 single qubit gates.

\begin{table}

\caption{Resource utilization summary for the various designed quantum circuits
in terms of number of gates used. \label{tab:Resource-utilization}}

\begin{centering}
\global\long\def\arraystretch{1.5}%
\begin{tabular}{|>{\centering}p{1.2cm}|>{\centering}p{1cm}|>{\centering}p{1cm}|>{\centering}p{1cm}|>
{\centering}p{1cm}|>{\centering}p{1cm}|}
\hline 
Parameters & Initial five qubit encoder & Initial five-qubit encoder NNC circuit & Optmized 5-qubit encoder & Five-qubit NNC circuit & 5-qubit encoder (\cite{gong2022experimental}
\tabularnewline
\hline 
\hline 
$H$ gate & 4 & 4 & 4 & 4 & 4 \tabularnewline
\hline 
$Z$ gate & 2 & 2 & 0 & 0 & 2 \tabularnewline
\hline 
$S$ gate & 0 & 0 & 0 & 0 & 3\tabularnewline
\hline
$S^{\dagger}$ gate & 0 & 0 & 0 & 0 & 3\tabularnewline
\hline
CNOT & 4 & 13 & 8 & 11 & 6\tabularnewline
\hline 
CZ & 6 & 6 & 0 & 0 & 0\tabularnewline
\hline 
\end{tabular}
\par\end{centering}
\end{table}

\section{Conclusions}

We present an optimized 5-qubit encoder circuit using 8 CNOT gates and 4 $H$ gates. It uses only one type of 2-qubit gate, i.e., the CNOT gates, which is more useful for practical applications. Further, it uses less number of 2-qubit gates, and is thus less prone to noise. 
%Future research should be directed towards of optimization of quantum circuits for longer codes such as 13-quibit code.
%However, it is difficult to say whether the circuit can be optimized further. Further research needs to be done towards automated tools for optimization of circuits.

\bibliographystyle{IEEEtran}
\bibliography{references}

\end{document}